\documentclass[journal]{IEEEtran}
\usepackage[cmex10]{amsmath}
\usepackage{amssymb}
\usepackage{graphicx}
\usepackage{amsthm}
\usepackage{setspace}
\usepackage{color}
\usepackage[justification=centering]{caption}
\usepackage{hyperref}
\usepackage{subfigure}
\usepackage{breakurl}
\usepackage[ruled]{algorithm2e}
\usepackage{algpseudocode}
\usepackage{amsmath}

\usepackage{graphics}
\usepackage{extarrows}
\usepackage{epsfig}
\newtheorem{thm}{\textit{Theorem}}
\ifCLASSINFOpdf
\else
\fi
\begin{document}
%
\title{Omnidirectional Quasi-Orthogonal Space-Time Block Coded Massive MIMO Systems }
%
%
%

\author{Can Liu, Xiang-Gen Xia, \emph{Fellow, IEEE}, Yongzhao Li, \emph{Senior Member, IEEE}, \\
Xiqi Gao, \emph{Fellow, IEEE},  and Hailin Zhang, \emph{Member, IEEE}
\thanks{C. Liu, Y. Li, and H. Zhang are with the State Key Laboratory of
Integrated Services Networks, Xidian University, Xi’an 710071, China
(e-mail: canliu@stu.xidian.edu.cn; yzhli@xidian.edu.cn; hlzhang@xidian.edu.cn).}
\thanks{X.-G. Xia is with the Department of Electrical and Computer Engineering,
University of Delaware, Newark, DE 19716 USA (e-mail: xxia@ee.udel.edu).}
\thanks{X. Q. Gao is with the National Mobile Communications Research Laboratory, Southeast University, Nanjing 210096, China (e-mail: xqgao@seu.edu.cn). \emph{Y. Li is the corresponding author.}}}

\maketitle

\vspace{-10pt}
\begin{abstract}
Common signals in public channels of cellular
systems are usually transmitted omnidirectionally from the base
station (BS). In recent years, both discrete and consecutive omnidirectional space-time block codings (STBC) have been proposed for massive multiple-input multiple-output (MIMO) systems with a uniform linear array (ULA) configuration to ensure cell-wide coverage. In these systems, constant received signal power at discrete angles, or constant received signal sum power in a few consecutive time slots at any angle is achieved. In addition, equal-power transmission per antenna and full spatial diversity can be achieved as well. {In this letter,  by utilizing the property of orthogonal complementary codes (OCCs), a new consecutive omnidirectional quasi-orthogonal STBC (QOSTBC) design is proposed, in which constant received sum power at any angle can be realized with equal-power transmission per antenna through one STBC transmission, and a higher diversity order of 4 can be achieved.  In addition, the proposed design can be further extended to the uniform planar array (UPA) configuration with the two-dimensional OCCs.}
\end{abstract}


%
\IEEEpeerreviewmaketitle

\section{Introduction}
\IEEEPARstart{T}{o} meet the challenging capacity requirement of the fifth generation (5G), massive multiple-input multiple-output (MIMO) system with tens to hundreds of antennas deployed at base station (BS) has attracted substantial attentions \cite{5G}. Public channels play important roles since
many essential common signals are provided to
users from BS through public channels. 

In order to broadcast common information from BS, discrete and consecutive omnidirectional space-time codings have been
recently proposed in \cite{Omni1}-\cite{constants} {for massive MIMO systems with a uniform linear array (ULA) configuration}. In  \cite{Omni1}\cite{Omni2}, by utilizing
the Zadoff-Chu (ZC) sequences, equal-power transmission per antenna and constant received signal power at finite discrete angles are satisfied, where the number of discrete angles is the same as the
number of transmit antennas. In \cite{constants}, a design of consecutive omnidirectional
space-time coding is proposed, where the orthogonal space-time block code (OSTBC),  Alamouti code (AC),  is used for 2 data streams, the sum of received signal powers at 2 consecutive time slots is constant at any angle, and equal-power transmission per antenna and diversity order of 2 are achieved as well. However, the design in \cite{constants} can only be applied  to AC. Although quasi-OSTBCs (QOSTBC) for 4 data streams of diversity order 4 are designed in \cite{Omni2}, {constant received signal power can be achieved only at finite discrete angles}. 

{In this letter, to further increase the diversity order over the AC coding in \cite{constants}, by utilizing the orthogonal complementary codes (OCCs) \cite{com1}, a new consecutive omnidirectional QOSTBC design is proposed, where the received signal sum power in 4 consecutive time slots at any angle is constant, and equal-power transmission per antenna at any time and full diversity order of 4 are achieved as well. {We want to emphasize that these three properties are new and additional to all the existing properties of the QOSTBC studies in, for example, \cite{qostbc}-\cite{signal2}. }
Unlike the discrete omnidirectional STBCs, the proposed design is insensitive to the number of BS antennas. Moreover, constructed with binary OCCs, high-resolution phase shifters are not necessary at the BS when employing the proposed STBCs, which will significantly reduce the energy consumption and the BS deployment expense. In addition, by utilizing the two-dimensional OCCs (2D-OCCs) \cite{com2}, similar omnidirectional STBCs can be designed for massive MIMO with uniform planar arrays (UPAs) equipped in BSs. }
%
%
%
%

%
%
%

 \vspace{-5pt}
\section{Problem Description}
 \vspace{-5pt}
\subsection{System Model}
In this letter, we consider STBC transmission for common information broadcasting. {For simplicity, we first consider that a BS is equipped with a ULA of $M$ antennas and serves $K$
users each with a single antenna.} The common information is mapped to an STBC ${{\bf{S}}\in\mathbb{C}^{M\times T}}$ with $M \ge T$ and transmitted from $M$ antennas of BS within $T$ time slots to all the users. The received signal of user $k$ can be written as
\begin{equation}
\left[ {{y_{k,1}},{y_{k,2}}, \ldots ,{y_{k,T}}} \right] = \sqrt {P_t} {{\bf{h}}_k^\mathbb{T}}{\bf{S}} + \left[ {{z_{k,1}},{z_{k,2}}, \ldots ,{z_{k,T}}} \right]
\end{equation}
where $P_t$ denotes the total transmit power, the channel ${\bf{h}}_k\in\mathbb{C}^{M\times 1}$ is assumed to keep constant within $T$ time slots, $\left(  \cdot  \right)^\mathbb{T}$ stands for the transpose, and ${z_{k,t}} \sim \mathcal{CN}\left( {0,\sigma _n^2} \right)$ is the additive white Gaussian noise (AWGN) at time slot $t$ $(t=1,2,\ldots, T)$. {Note that, different from the multi-user MIMO downlink transmissions where the BS transmits different signals to different users and $K$ is upper bounded, in this letter, since the BS transmits the same common signals to all the users in public channels, there is no interference between different users. Therefore, $K$ can be arbitrary here.}

To decode the transmitted information symbols in codeword  $\bf{S}$, the instantaneous CSI ${\bf{h}}_k$ must be known at the user side. However, in massive MIMO systems, since the number of BS antennas is very large and the number of downlink resources needed for pilots is  proportional to the number of BS antennas, the downlink channel estimation becomes challenging. In order to reduce the pilot overhead, a dimensional-reduced STBC is utilized in, for example, \cite{Omni1}-\cite{constants}, where a high dimensional STBC is composed by a precoding matrix ${\bf{W}}$ and a low dimensional STBC ${\bf{X}}$, i.e., $\bf{S}=\bf{WX}$, where ${\bf{W}}\in\mathbb{C}^{M\times N}$ is a tall precoding matrix independent of the channel or the information data {(since users may be inactive and no feedback is available)}, and ${\bf{X}}\in\mathbb{C}^{N\times T}$ is a low dimensional STBC modulated by the common information data. To decode the common information data, users only need to estimate the effective channel ${\bf{W}}{^\mathbb{T}\bf{h}}_k$ of dimension $N \times 1$ with $N\ll M$ instead. To normalize the total average transmission power at the BS side, we assume that $\mathbb{E}\left( {{\bf{X}}{{\bf{X}}^H}} \right) = {T \mathord{\left/
 {\vphantom {T N}} \right.
 \kern-\nulldelimiterspace} N} \cdot {{\bf{I}}_N}$ and ${\rm{tr}}({\bf{WW}}^H)=\emph{N}$ { where $\mathbb{E}\left(\cdot\right)$ denotes the expectation, $\left(\cdot\right)^H$ donotes the Hermitian operation, ${\bf{I}}_N$ is the  $N \times N$ identity matrix and $\rm{tr}\left(  \cdot  \right)$ stands for the trace of a matrix}.
 \vspace{-8pt}
\subsection{Criteria of Consecutive Omnidirectional STBC}
For consecutive omnidirectional STBC design, the following three criteria should be guaranteed \cite{constants}.

{\bf{1. Criterion of Constant Instant Transmission Power at
Each Antenna at Any Time Slot $t$}}:  Assume ${\bf{x}}_t$ is the $t$-th column vector of the low dimensional STBC ${\bf{X}}$, then ${\bf{Wx}}_t$ is the transmitted signal in BS at time slot $t$. To sufficiently utilize all the power amplifier (PA) capacities of BS antennas, the precoding design should satisfy
\begin{equation}
\left| {{{\left[ {{\bf{W}}{{\bf{x}}_t}} \right]}_m}} \right| = \frac{1}{{\sqrt M }},\;m = 1, \ldots M\label{c4}
\end{equation}
at any time slot $t$, where ${{{\left[ {{\bf{W}}{{\bf{x}}_t}} \right]}_m}}$ denotes the transmitted signal
on the $m-$th antenna at time slot $t$.

{\bf{2. Criterion of Full Diversity Order $T$}}: The STBC ${\bf{S}}={\bf{WX}}$ satisfies the
full column rank, i.e., for any two distinct STBC codewords
${\bf{S}}_1$ and ${\bf{S}}_2$ of ${\bf{S}}$, the difference $ {\bf{S}}_1-{\bf{S}}_2$ has full column rank $T$.

{\bf{3. Criterion of Constant Received Signal Sum Power at Any
Angle}}: The corresponding transmitted signal in the angle domain {under a ULA configuration} can be written as
\begin{equation}
S_t(\omega) = {{\bf{a}}(\omega)} \cdot {\left[ {{\bf{W}}{{\bf{x}}_t}} \right]} , \label{12}
\end{equation}
where ${{\bf{a}}(\omega)} = \left[ {1,{e^{-{\rm{j}}{\omega}}} ,\cdots ,{e^{-{\rm{j}}(M - 1){\omega}}}} \right]$ is the antenna array response vector under the ULA setup and $\omega = 2\pi \frac{{{d}}}{\lambda }\sin \left( {{\theta }} \right)$ with carrier
wavelength $\lambda$, antenna spacing $d$ and azimuth angle of
departure (AoD) $\theta$. The criterion is
\begin{equation}
\sum\limits_{t = 1}^T {|S_t(\omega){|^2}}  = c,\,\forall {\theta} \in (-\pi,\pi ]
\label{16}
\end{equation}
for a positive constant $c$. {For a UPA configuration in BSs,
the criterion can be rewritten as
\begin{equation}
\sum\limits_{t = 1}^T {|S_t(\omega,\upsilon){|^2}}  = c,\,\forall {\theta} \in (-\pi,\pi ], {\phi}\in [0,\pi]
\label{upa}
\end{equation}
for a positive constant $c$. Here, $\omega = 2\pi \frac{{{d_a}}}{\lambda }\sin \left( {{\theta }}\right)\sin \left( {{\phi }} \right)$, and  $\upsilon=2\pi \frac{{{d_e}}}{\lambda }\cos \left( {{\phi }} \right)$ with antenna spacings $d_a$ and $d_e$, and AoDs $\theta$ and $\phi$ in azimuth and elevation, respectively.}

From \eqref{12}, one can see that, $S_t(\omega)$ is the Fourier transform (FT) of ${{\bf{W}}{{\bf{x}}_t}}$. 
Note that the received signal power of $S_t(\omega)$ at time slot $t$ cannot be constant for any angle $\theta$ as mentioned in \cite{constants}. It is the motivation in \cite{constants} to consider a sum of a few consecutive signal powers as \eqref{16}. 

In this letter, we design precoded QOSTBC for { $N=4$} data streams, {i.e., symbol rate 1, }with spatial diversity order of 4. Note that OSTBCs for $T=N=4$ can accommodate at most 3 data streams \cite{ostbc}, {i.e., their symbol rates are at most 3/4, }although they can achieve the spatial diversity order of 4.
 \vspace{-8pt}
\subsection{Orthogonal Complementary Codes}
Some useful mathematical results are reviewed here to help the omnidirectional STBC design. Let 
$\left\{ {{{\bf{c}}_{i,j}},1 \le i \le p,1 \le j \le q} \right\}$ be a set of orthogonal complementary codes, where each code ${{\bf{c}}_{i,j}}$ is a sequence \footnote{{In \cite{constants}, complementary pair is introduced to construct the precoding matrix ${\bf{W}}$ for AC. However, it does not work for QOSTBC due to the nonorthogonality. Therefore, OCC is introduced here to overcome this problem, and more details are given in Section III.}}
of length $L$. Every $i$-th subset $\left\{ {{\bf{c}}_{i,1}}, {{\bf{c}}_{i,2}}, \ldots, {{\bf{c}}_{i,q}}\right\}$ is a complementary set of $q$ sequences, and for $i\neq i'$, the $i$-th and $i'$-th subsets are mutually orthogonal complementary \cite{com1}\cite{com2}. The OCC has the following properties.
\begin{equation}
\left\{ \begin{array}{l}
\sum\limits_{j= 1}^q {{R _{{{\bf{c}}_{i,j}}}}\left( {\tau} \right)}  = 0,\;\forall \tau \ne 0,1 \le i \le p\\
\sum\limits_{j = 1}^q {{R _{{{\bf{c}}_{i,j}}{{\bf{c}}_{i',j}}}}\left( {\tau} \right)}  = 0,\; \forall \tau,1 \le i \ne i' \le p
\end{array} \right.\label{occ}
\end{equation}
with
\begin{equation}
\left\{ \begin{array}{l}
{R _{{{\bf{c}}_{i,j}}}}\left( {\tau} \right) = \sum\limits_{l = 1}^{{L}} {{{\left[ {{{\bf{c}}_{i,j}}} \right]}_{l}}{{\left[ {{{\bf{c}}_{i,j}}} \right]}_{l + \tau}^*}}  \\
{R _{{{\bf{c}}_{i,j}}{{\bf{c}}_{i',j}}}}\left( {\tau} \right) = \sum\limits_{l = 1}^{{L}} {{{\left[ {{{\bf{c}}_{i,j}}} \right]}_{l}}{{\left[ {{{\bf{c}}_{i',j}}} \right]}_{l + \tau}^*}} 
\end{array} \right. \label{acf}
\end{equation}
where ${{R _{{{\bf{c}}_{i,j}}}}\left( {\tau} \right)}$ and ${{R _{{{\bf{c}}_{i,j}}{{\bf{c}}_{i',j}}}}\left( {\tau} \right)}$ are the aperiodic autocorrelation function (AACF) and the aperiodic crosscorrelation function (ACCF) for shift $\tau$, respectively, $\left( \cdot\right)^*$ denotes the complex conjugate, and ${{\left[ {{{\bf{c}}_{i,j}}} \right]}_{l}}$ is the $l$-th element of ${\bf{c}}_{i,j}$ if $1 \le l \le L$, and 0 otherwise. 

%
%
 \vspace{-3pt} 
\section{ Omnidirectional QOSTBC Design for A ULA Configuration}
 \vspace{-3pt}

OSTBCs have both advantages of complex symbol-wise maximum-likelihood (ML) decoding and full diversity. However,
their symbol rates are upper bounded by 3/4 for more than two
antennas for complex symbols \cite{ostbc} as mentioned earlier. Therefore, QOSTBCs are proposed in \cite{qostbc}\cite{minimal} where the orthogonality is relaxed
to achieve high symbol transmission rate but with a more complex symbol
pair-wise ML decoding.  By rotating the constellations of the complex symbols, the QOSTBCs can further achieve full diversity \cite{signal}-\cite{signal1}. We next want to study the consecutive omnidirectional STBC design based on the QOSTBCs and OCCs for ULA configuration.

Consider the low dimensional QOSTBC of Tirkkonen, Boariu, and Hottinen (TBH) scheme \cite{minimal} as an example, and constellation rotations \cite{signal1} are applied to achieve the diversity order of $T=4$, 
\begin{equation}
{\bf{X}}={\bf{X}}_Q \buildrel \Delta \over =  \left[ {\begin{array}{*{20}{c}}
{{x_1}}&{x_2^*}&{{x_3}}&{x_4^*}\\
{{x_2}}&{ - x_1^*}&{{x_4}}&{ - x_3^*}\\
{{x_3}}&{x_4^*}&{{x_1}}&{x_2^*}\\
{{x_4}}&{ - x_3^*}&{{x_2}}&{ - x_1^*}
\end{array}} \right] \label{qo}
\end{equation}
where $x_1$, $x_2$ are taken from a symbol constellation $\mathcal{S}$, and $x_3$, $x_4$ are taken from the rotated symbol constellation $e^{{\rm{j}}\varphi}\mathcal{S}$.
The correlation matrix of $ {\bf{X}}_Q$ in  \eqref{qo} is
\begin{equation}
{{\bf{X}}_Q}{\bf{X}}_Q^H = \left[ {\begin{array}{*{20}{c}}
\alpha &0&\beta &0\\
0&\alpha &0&\beta \\
\beta &0&\alpha &0\\
0&\beta &0&\alpha
\end{array}} \right] = \alpha {{\bf{I}}_4} + \beta {\bf{{\Pi}}}_2
\end{equation}
where $\alpha  = \sum\limits_{i = 1}^4 {{{\left| {{x_i}} \right|}^2}} $, $\beta  = 2{\mathop{\rm Re}\nolimits} \left( {{x_1}x_3^* + {x_2}x_4^*} \right)$, ${\mathop{\rm Re}\nolimits} \left(  \cdot  \right)$ denotes the real part of a
complex number, and ${\bf{\Pi}} _2 = \left[ {\begin{array}{*{20}{c}}
{\bf{0}}&{{{\bf{I}}_2}}\\
{{{\bf{I}}_2}}&{\bf{0}}
\end{array}} \right]$
where ${\bf{I}}_2$ is the  $2 \times 2$ identity matrix. Note that while $x_1,x_2,x_3,x_4$ are independent with each other, $\beta$ is not always possible to be 0. 
Then, (\ref{16}) can be rewritten as
\begin{equation}
\small
\hspace{-0.3in}
\begin{array}{l}
\begin{aligned}
\!&\sum\limits_{t = 1}^4 {|{S_t}(\omega ){|^2}}  
= {{\bf{a}}}(\omega ){\bf{W}}{{\bf{X}}_Q}{\bf{X}}{_Q^H}{{\bf{W}}^H}{\bf{a}}^H(\omega )\\
 = &\alpha {{\bf{a}}}(\omega ){\bf{W}}{{\bf{W}}^H}{\bf{a}}^H(\omega ) + \beta {{\bf{a}}}(\omega ){\bf{W}}\Pi _2{{\bf{W}}^H}{\bf{a}}^H(\omega )\\

= &\alpha \sum\limits_{n = 1}^4 {{{\bf{a}}}(\omega ){{\bf{w}}_n}{\bf{w}}_n^H{\bf{a}}^H(\omega )}  \\

 + &\beta \sum\limits_{n = 1}^2 {{{\bf{a}}}(\omega )\left( {{{\bf{w}}_n}{\bf{w}}_{n + 2}^H + {{\bf{w}}_{n + 2}}{\bf{w}}_n^H} \right){\bf{a}}^H(\omega )} \\

 = &\!\alpha\! \sum\limits_{n = 1}^4 \!{|{W_n}(\omega ){|^2}}  \!\!+ \!2\beta {\mathop{\rm Re}\nolimits} \left(\sum\limits_{n = 1}^2\! {{W_n}\!(\omega ){W_{n + 2}^*}(\omega )}\right)
\end{aligned}
\end{array}\label{sum}
\setlength{\abovedisplayskip}{2.5pt}
\setlength{\belowdisplayskip}{2.5pt}
\end{equation}
where ${\bf{w}}_n$ is the $n$-th column vector of ${\bf{W}}$, and ${W_n}(\omega )$ is the FT of ${\bf{w}}_n$ for $n=1,\cdots,4$. 
\vspace{-5pt}
\begin{thm}\label{lemma1}
If the sequence sets $\left\{{\bf{w}}_1,{\bf{w}}_2\right\}$ and $\left\{{\bf{w}}_3,{\bf{w}}_4\right\}$ are mutually orthogonal complementary, and $\alpha$ is constant for signal constellation ${\mathcal{S}}$, then, the criterion 3 {for a ULA configuration,} i.e., \eqref{16}, of constant received signal sum power with $T=4$ holds { for all $\theta \in (-\pi,\pi]$}. 
\end{thm}
\vspace{-9pt}
\begin{proof}
See Appendix A.
\end{proof}
\vspace{-3pt}

In addition, the QOSTBC design should satisfy the criterion of equal power at
each antenna as well. Consequently, we propose the precoding matrix ${\bf{W}}={\bf{W}}_Q$ as follows. 

{Consider a ULA is equipped in the BS,} we assume that the number of BS  antennas $M$ is an integer multiple of 4, and let $\left\{{\bf{c}}_1, \ldots, {\bf{c}}_4 \right\}$ be a set of binary OCCs of length { $L=M/4$}, in which $\left\{{\bf{c}}_1,{\bf{c}}_2\right\}$ and $\left\{{\bf{c}}_3,{\bf{c}}_4\right\}$ are two sets of complementary pairs of components either 1 or -1 and they are mutually orthogonal complementary. The precoding matrix ${\bf{W}}_Q$ is
\begin{equation}
{\bf{W}}_Q =\sqrt{\frac{4}{ M }}
[  {{{\bf{c}}_1} \otimes {{\bf{u}}_1}},{{{\bf{c}}_2} \otimes {{\bf{u}}_2}} ,
 {{{\bf{c}}_3} \otimes {{\bf{u}}_3}} ,{{{\bf{c}}_4} \otimes {{\bf{u}}_4}} ]
\label{qqostbc}
\end{equation}
where
${{\bf{u}}_i}$ is the $i$-th column vector of the $4\times 4$ identity matrix ${{\bf{I}}_4}$, and $\otimes$ denotes the Kronecker product. Clearly, ${\bf{W}}_Q$ has rank 4, i.e., it has full column rank. The signal constellation $\mathcal{S}$ is selected to be a phase shift keying
(PSK), i.e., $x_i \in\mathcal{S}_{PSK}={\frac{1}{2}}\left\{ {1,{e^{{\rm{j}}2\pi /\vartheta }}, \ldots ,{e^{{\rm{j}}2\pi \left( {\vartheta  - 1} \right)/\vartheta }}} \right\}$ for some positive integer $\vartheta$.  The optimal rotation angle for PSK signal $x_3$ and $x_4$ is ${\pi  \mathord{\left/
 {\vphantom {\pi  \vartheta }} \right.
 \kern-\nulldelimiterspace} \vartheta }$ when $\vartheta$ is even and ${\pi  \mathord{\left/
 {\vphantom {\pi  (2\vartheta) }} \right.
 \kern-\nulldelimiterspace} (2\vartheta) }$ or ${3\pi  \mathord{\left/
 {\vphantom {3\pi  (2\vartheta) }} \right.
 \kern-\nulldelimiterspace} (2\vartheta) }$ when $\vartheta$ is odd \cite{signal1}. In this way, the precoding design will satisfy all the three criteria as we shall see below. 

First, let $c_{n,i}$ be the $i$-th element of ${\bf{c}}_n$, we have
\begin{equation}
\begin{array}{l}

\!\!{\bf{S}} = {{\bf{W}}_Q}{{\bf{X}}_Q}\\
\!\!= \!\!\!\sqrt{\frac{4}{ M }}\!{\begin{bmatrix}
{{c_{1,1}}{x_1}}&{{c_{1,1}}x_2^*}&{{c_{1,1}}{x_3}}&{{c_{1,1}}x_4^*}\\
{{c_{2,1}}{x_2}}&{ - {c_{2,1}}x_1^*}&{{c_{2,1}}{x_4}}&{ - {c_{2,1}}x_3^*}\\
 \vdots & \vdots & \vdots & \vdots \\
{{c_{3,{L}}}{x_3}}&{{c_{3,{L}}}x_4^*}&{{c_{3,{L}}}{x_1}}&{{c_{3,{L}}}x_2^*}\\
{{c_{4,{L}}}{x_4}}&{ - {c_{4,{L}}}x_3^*}&{{c_{4,{L}}}{x_2}}&{ - {c_{4,{L}}}x_1^*}
\end{bmatrix}}.
\end{array}
\end{equation}
Since ${\bf{c}}_n$ is a sequence of 1's and -1's, and $x_i$ is constant even with the rotations, it is easy to see that, all the elements in ${\bf{S}}$ have the same amplitude, so the criterion 1 of equal-power transmission per antenna at any time holds.

Since the low dimensional QOSTBC ${\bf{X}}_Q$ in (\ref{qo}) has 
diversity order of 4 after the optimal angle rotation, and the precoding matrix ${\bf{W}}_Q$ is
a constant full column rank matrix, it is clear that the low dimensional QOSTBC ${\bf{S}}={\bf{W}}_Q{\bf{X}}_Q$ has
diversity order $T = 4$, i.e., the criterion 2 of full diversity order holds. 

Then, let us prove that it
satisfies the constant received signal sum power criterion 3. When PSK signals are adopted, we have $\alpha= \sum\limits_{i = 1}^4 {{{\left| {{x_i}} \right|}^2}}=1$. Let $C_n(\omega)$ be the FT of ${{\bf{c}}_n}$, so ${W_n}\left( \omega  \right) = \sqrt{\frac{4}{ M }}{e^{ - {\rm{j}}\left( {n - 1} \right)\omega }}{C_n}(4\omega )$. According to (\ref{sum}) and Theorem \ref{lemma1}, we have
\begin{equation}
\small
\begin{array}{l}
\begin{aligned}
\!\!\sum\limits_{t = 1}^4& {|{S_t}(\omega ){|^2}}  = \frac{{4\alpha }}{M}\sum\limits_{n = 1}^4 {|{C_n}(4\omega ){|^2}} \\
 &+ \frac{{8\beta }}{M}{\mathop{\rm Re}\nolimits} \left( {{e^{  2{\rm{j}}\omega }}\sum\limits_{n = 1}^2 {{C_n}(4\omega ){C_{n + 2}^*}(4\omega )} } \right) = 4
\end{aligned}
\end{array}
\setlength{\abovedisplayskip}{2pt}
\setlength{\belowdisplayskip}{2pt}
\end{equation}
 which is constant for $\theta \in (-\pi,\pi]$ {under a ULA configuration}.
 
In summary, the above precoded QOSTBC satisfies all the three criteria presented in Section II. Similarly, the omnidirectional QOSTBC based on Jafarkhani scheme \cite{qostbc} or other schemes can be done and we omit the details here.
In addition, to decrease the ML decoding complexity, we also find that the proposed scheme can work for the QOSTBC with minimum decoding complexity (MDC) \cite{mdc}\cite{signal2} when QPSK modulation ($x_i \in {\frac{1}{2}}\left\{ \pm 1, \pm j \right\}$) is used (refer to the Case 2 and (42) in \cite{signal2}).
Instead of complex symbol pair-wise ML decoding, QOSTBC with MDC
has real symbol pair-wise ML decoding which has the same complexity as
complex symbol-wise decoding.


Note that the length of a binary OCC may not be arbitrary. The construction of binary OCC is possible with any length $L = {2^a} \cdot {10^b} \cdot {26^c}$ for all integers $a$, $b$, $c\ge 0$, and there is no binary OCC containing two sequences whose length cannot be expressed as a sum of two squares \cite{com1}\cite{com2}. Therefore, the number of BS antennas may not be arbitrary either for our proposed designs in this letter. 
\section{Omnidirectional QOSTBC Design for A UPA Configuration}
For massive MIMO systems under a UPA configuration, similar omnidirectional QOSTBC designs can be done by utilizing the property of 2D-OCCs \cite{com2}.
\subsection{Two-Dimensional Orthogonal Complementary Codes}
We denote $\left\{ {{{\mathbf{B}}_{i,j}},1 \le i \le P,1 \le j \le Q} \right\}$~as a set of 2D-OCCs, where each element ${{\mathbf{B}}_{i,j}}$ is an $L_1 \times L_2$ matrix. Every $i$-th subset $\left\{ {{\mathbf{B}}_{i,1}}, {{\mathbf{B}}_{i,2}}, \ldots, {{\mathbf{B}}_{i,Q}}\right\}$ is a 2D complementary set of $Q$ matrice. For $i\neq k$, the $i$-th and $k$-th subsets are mutually orthogonal complementary \cite{com2}.  The 2D-OCC has the following properties
\begin{equation}
\!\!\left\{ \begin{array}{l}
\!\sum\limits_{j = 1}^Q {{R _{{{\mathbf{B}}_{i,j}}}}\left( {m,n} \right)} \! = \!0,\;\forall \left( {m,n} \right) \!\ne\! \left( {0,0} \right),1 \!\le\! i \le P\\
\!\sum\limits_{j = 1}^Q {{R _{{{\mathbf{B}}_{i,j}}{{\mathbf{B}}_{k,j}}}}\left( {m,n} \right)}  \!= \!0,\; \forall \left( {m,n} \right),1 \!\le \! i \!\ne\! k \le P
\end{array} \right.\label{occ}
\end{equation}
with
\begin{equation}
\!\!\!\left\{ \begin{array}{l}
\!\!\!\!{R _{{{\mathbf{B}}_{i,j}}}}\left( {m,n} \right) = \sum\limits_{p = 1}^{{L_1}} {\sum\limits_{q = 1}^{{L_2}} {{{\left[ {{{\mathbf{B}}_{i,j}}} \right]}_{p,q}}{{\left[ {{{\mathbf{B}}_{i,j}}} \right]}_{p + m,q + n}^*}} } \\
\!\!{R _{{{\mathbf{B}}_{i,j}}{{\mathbf{B}}_{k,j}}}}\left( {m,n} \right) = \sum\limits_{p = 1}^{{L_1}} {\sum\limits_{q = 1}^{{L_2}} {{{\left[ {{{\mathbf{B}}_{i,j}}} \right]}_{p,q}}{{\left[ {{{\mathbf{B}}_{k,j}}} \right]}_{p + m,q + n}^*}} }
\end{array} \right. \label{acf}
\end{equation}
where ${{R _{{{\mathbf{B}}_{i,j}}}}\left( {m,n} \right)}$ and ${{R _{{{\mathbf{B}}_{i,j}}{{\mathbf{B}}_{k,j}}}}\left( {m,n} \right)}$ are the 2D aperiodic autocorrelation function (AACF) and  2D aperiodic crosscorrelation function (ACCF) for vertical shift $m$ and horizontal shift $n$, respectively. Here, if $1 \le p \le L_1$ and $1 \le q \le L_2$, ${{\left[ {{{\mathbf{B}}_{i,j}}} \right]}_{p,q}}$ is the $(p,q)$-th element of ${\mathbf{B}}_{i,j}$, otherwise ${{\left[ {{{\mathbf{B}}_{i,j}}} \right]}_{p,q}}=0$. 
\begin{table}
\caption{Binary OCC of length 16}
\setlength{\tabcolsep}{1mm}{
\begin{tabular}{c|cccccccccccccccc}
\hline
${\bf{c}}_1$ &1&1&1&-1&1&1&-1&1&1&1&1&-1&-1&-1&1&-1\\
${\bf{c}}_2$ &1&1&1&-1&1&1&-1&1&-1&-1&-1&1&1&1&-1&1\\
${\bf{c}}_3$ &1&-1&1&1&1&-1&-1&-1&1&-1&1&1&-1&1&1&1\\
${\bf{c}}_4$ &1&-1&1&1&1&-1&-1&-1&-1&1&-1&-1&1&-1&-1&-1\\
\hline
\end{tabular}}
\vspace{-10pt}
\end{table}
\begin{figure}
  \centering
  \setlength{\abovecaptionskip}{-0cm}
\setlength{\belowcaptionskip}{-0.5cm}
  \includegraphics[width=.65\linewidth]{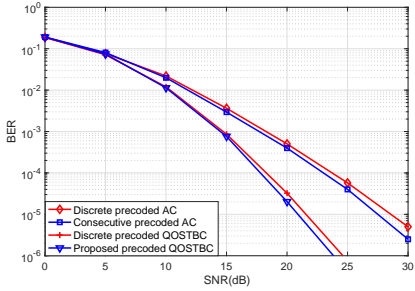}
  \caption{{BER performance versus SNR for 2 bps/Hz.}}
\end{figure}

\subsection{Omnidirectional QOSTBC Design}
We consider the BS under a UPA configuration of $M=N_a\times N_e$ antennas, and there are  $N_a$ antennas in each row and $N_e$ antennas in each column. Similar to the design for ULA in \eqref{qqostbc}, when the QOSTBC of TBH scheme \cite{minimal} in (\ref{qo}) is adopted as the low dimensional STBC, we can construct the precoding matrix ${\mathbf{W}}_Q$ based on 2D-OCCs as
\begin{equation}
\small
{\mathbf{W}}_{\!Q}\! =\!\!\sqrt{\!\frac{4}{ M }}
[ {\textrm{vec}}\!\left(\! {{{\mathbf{B}}_1} \!\!\otimes\! {{\mathbf{U}}_1}}\! \right)\!,{\textrm{vec}}\!\left(\! {{{\mathbf{B}}_2}\!\! \otimes\! {{\mathbf{U}}_2}} \!\right)\!,
 {\textrm{vec}}\!\left(\! {{{\mathbf{B}}_3} \!\!\otimes\! {{\mathbf{U}}_3}} \!\right)\!,{\textrm{vec}}\!\left( {{{\mathbf{B}}_4} \!\!\otimes\! {{\mathbf{U}}_4}}\! \right)]
\label{ostbc}
\end{equation}
with $\left\{{\mathbf{B}}_1, \ldots, {\mathbf{B}}_4 \in\mathbb{C}^{N_e/2 \times N_a/2}\right\}$ as a set of 2D binary orthogonal complementary codes, in which $\left\{{\mathbf{B}}_1,{\mathbf{B}}_2\right\}$ and $\left\{{\mathbf{B}}_3,{\mathbf{B}}_4\right\}$ are two sets of 2D complementary pairs of components either 1 or -1 and they are mutually orthogonal complementary. Here, 
${{\mathbf{U}}_1} \!\!=\!\!\! \left(\! \!{\begin{array}{*{10}{c}}
1\!&0\\
0\!&0
\end{array}}\!\! \right)\!$, ${{\mathbf{U}}_2} \!\!= \!\!\left(\!\! {\begin{array}{*{20}{c}}
0\!&0\\
1\!&0
\end{array}} \!\!\right)$, ${{\mathbf{U}}_3} \!\!= \!\!\left(\!\! {\begin{array}{*{20}{c}}
0\!&1\\
0\!&0
\end{array}} \!\!\right)$, and ${{\mathbf{U}}_4} \!\!=\!\! \left( \!\!{\begin{array}{*{20}{c}}
0\!&0\\
0\!&1
\end{array}} \!\!\right)$.

It is easy to see that, the criterion 1 of equal-power transmission per antenna at any time holds for this design. With the low dimensional QOSTBC ${\bf{X}}_Q$ after the optimal angle rotation, the criterion 2 of full diversity order holds as well. 

When PSK signals are adopted, we have $\alpha= \sum\limits_{i = 1}^4 {{{\left| {{x_i}} \right|}^2}}=1$. Let $\mathcal{B}_n(\omega,\upsilon )$ and ${W_n}\left( \omega, \upsilon \right)$ be the 2D FT of ${{\bf{B}}_n}$ and ${{{\mathbf{B}}_n} \otimes {{\mathbf{U}}_n}}$, so ${W_n}\left( \omega, \upsilon \right) = \sqrt{\frac{4}{ M }}{e^{ - {\rm{j}}\left( {p - 1} \right)\omega }}{e^{ - {\rm{j}}\left( {q - 1} \right)\upsilon }}{\mathcal{B}_n}(2\omega,2\upsilon )$, where $\left(p,q \right)$ is the position of the nonzero element in ${{\mathbf{U}}_i}$, i.e. ${\left[ {{{\mathbf{U}}_i}} \right]_{p,q}} = 1$. According to (\ref{sum}) and Theorem \ref{lemma1}, similar conclusion
\begin{equation}
\small
\begin{array}{l}
\begin{aligned}
\!\!\sum\limits_{t = 1}^4& {|{G_t}(\omega,\upsilon ){|^2}}  = \frac{{4\alpha }}{M}\sum\limits_{n = 1}^4 {|{\mathcal{B}_n}(2\omega,2\upsilon ){|^2}} \\
 &+ \frac{{8\beta }}{M}{\mathop{\rm Re}\nolimits} \left( {{e^{  {\rm{j}}\omega }}\sum\limits_{n = 1}^2 {{\mathcal{B}_n}(2\omega,2\upsilon ){\mathcal{B}_{n + 2}^*}(2\omega,2\upsilon )} } \right) = 4
\end{aligned}
\end{array}
\setlength{\abovedisplayskip}{2pt}
\setlength{\belowdisplayskip}{2pt}
\end{equation}
 can be achieved which is constant for all $\theta \in (-\pi,\pi]$ and $\phi\in[0,\pi]$.
 
 Therefore, the precoded QOSTBC for a UPA configuration satisfies all the three criteria presented in Section II as well.

\vspace{-5pt}
\section{Numerical Results}
\vspace{-2pt}
\begin{figure}
  \centering
   \setlength{\abovecaptionskip}{-0cm}
\setlength{\belowcaptionskip}{-0.5cm}
  \includegraphics[width=0.65\linewidth]{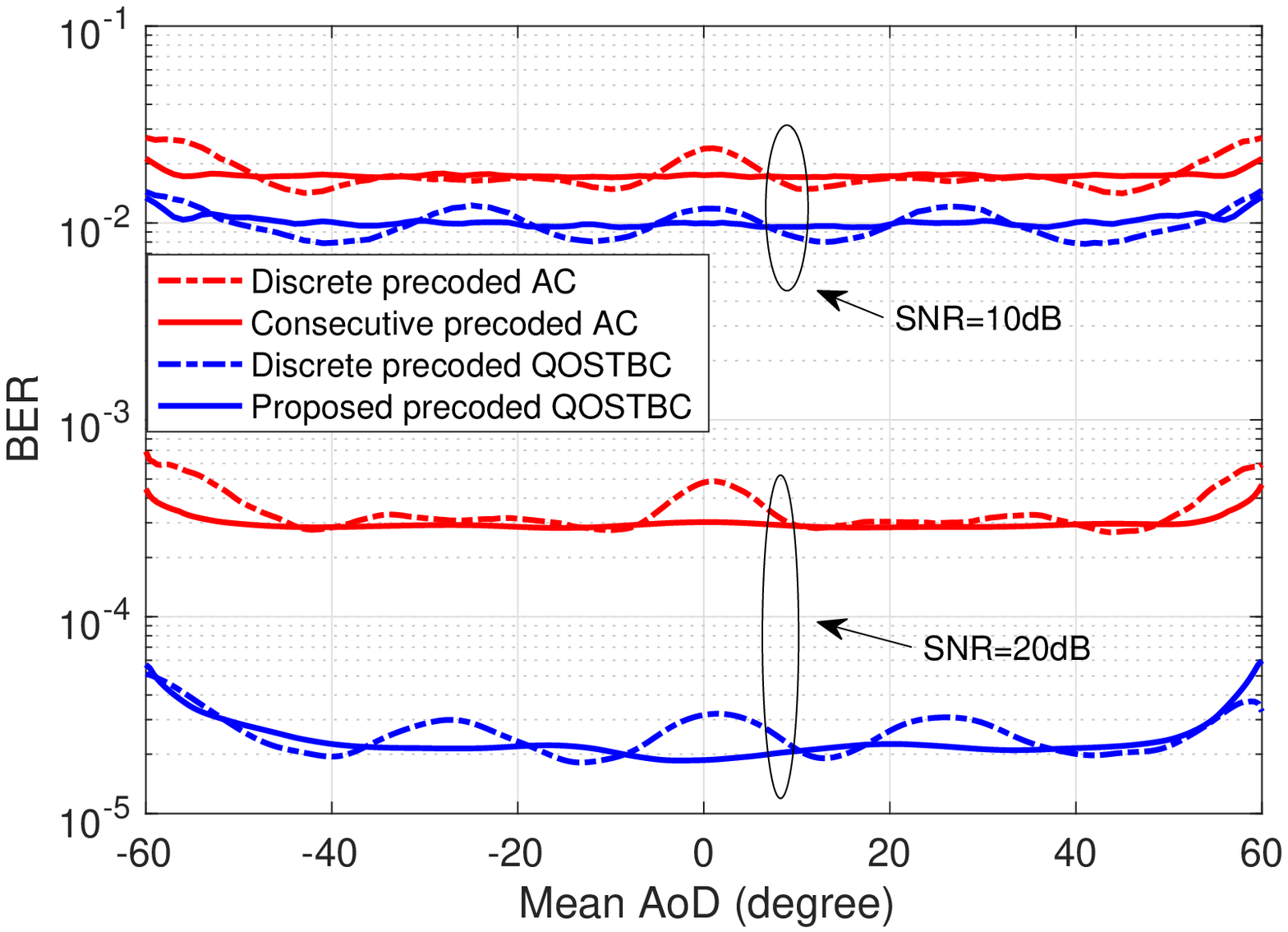}
  \caption{{ BER performance versus mean AoD for 2 bps/Hz.}}
\end{figure}

In this section, we will evaluate the performance of the proposed omnidirectional QOSTBC design. 
We first consider that the BS is a ULA of $M=64$ antennas in a $120^\circ$ sector, where the antenna space is $d=\lambda /\sqrt{3}$, and serves $K=300$ users each with a single antenna. 
The channel model here refers to the model in \cite{Omni2}. To represent that users have different angles with respect to the ULA of the BS within the sector, the mean AoD $\theta_0$ is randomly selected in uniform distribution on $\left[ -60^\circ, 60^\circ\right]$, while the angle spread (AS) $\sigma  = {5^\circ}$. 
The signal constellation in all the schemes is QPSK, while for either the discrete \cite{Omni2} or the preposed precoded QOSTBC, the optimal rotation angle of $\pi / 4$ is adopted \cite{signal1}. 
The binary OCC, which is shown in Table I \cite{com1}\cite{com2}, is used to obtain the proposed precoding matrix. 

First, we have the average bit error rate (BER) performance with respect to the signal-to-noise ratio (SNR) value, i.e., $P_t/\sigma_n^2$, as shown in Fig. 1. 
{ We can see that although all the schemes can achieve full diversity order of 2 or 4, consecutive precoded STBCs (either AC \cite{constants} or preposed QOSTBC) always outperform the discrete ones obtained in \cite{Omni2} with the same low dimensional STBCs in high SNR regime.}
This is because when the antenna number is limited, the angle resolution of the discrete precoded STBCs is not enough. 

Then, we evaluate the BER performance with respect to the mean AoD $\theta_0$ to verify the ability
of omnidirectional transmission for different STBC designs which is shown in Fig. 2.
{ We can see that compared with the discrete precoded STBCs, the consecutive precoded STBCs have flatter BER performance for different values of mean AoDs in both low and high SNR regimes, since the sum of the
received signal powers of 2 or 4 consecutive time slots in the  consecutive precoded STBCs is constant at any
angle, rather than just at finite discrete angles.}

Finally, we evaluate the beam pattern of the proposed omnidirectional STBC design based on 2D-OCCs.
Here, we consider that the BS is a UPA of $M=256$ antennas in a $120^\circ$ sector and $N_a=N_e=16$, where the antenna space is $d_a=d_e=\lambda /2$.The channel model here refers to the model in \cite{JSDM}. The 2D binary OCCs \cite{com1}\cite{com2}, which is shown in Fig. 3, are used to obtain the proposed precoding matrix. Let $x_1=1$, $x_2=\rm{j}$, $x_3=(1+\rm{j})/\sqrt{2}$, and $x_4=-1$. Fig. 4(e) shows the beam pattern of sum power in 4 consecutive time slots of one QOSTBC of TBH scheme, while Figs. 4(a-d) represent beam patterns of the first to the fourth time slots, respectively. It can be seen that, the consecutive omnidirectional STBC based on 2D-OCCs can realize omnidirectional transmission via a complete STBC transmission.
\begin{figure}
  \centering
  \includegraphics[width=0.7\linewidth]{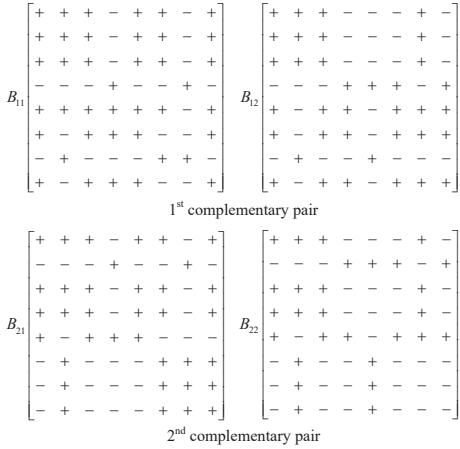}
  \caption{A set of 2D binary OCCs.}
  \label{omni11}
\end{figure}
 \begin{figure}
  \centering
  \includegraphics[width=0.9\linewidth]{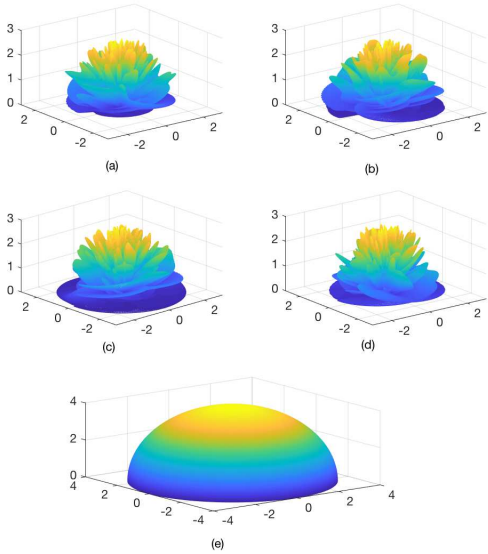}
  \caption{Beam pattern for $16\times 16$ UPA with consecutive omnidirectional transmission. (a) the first time slot, (b) the second time slot,  (c) the third time slot, (d) the fourth time slot, (e) sum power in 4 consecutive time slots.}
  \vspace{-20pt}
\end{figure}

\vspace{-7pt}
\section{Conclusion}
In this letter, consecutive omnidirectional QOSTBC designs for ULA and UPA configuration are proposed to guarantee omnidirectional transmission, i.e., the sum of $T=4$ consecutive received signal power is constant at any angle, equal instantaneous power on each transmit
antenna, and achieve the full diversity of
the low-dimensional QOSTBCs. 
Compared with the discrete omnidirectional precoding design, our proposed omnidirectional QOSTBCs for a ULA configuration has flatter BER performance for all angles of DoA. In addition,  the proposed omnidirectional QOSTBCs for a UPA configuration can realize omnidirectional transmission in any azimuth and elevation angles.


%
%
\appendices
\section{Proof of Theorem 1}\label{app1}
Let ${{\bf{w}}_n} = (w_{n,m})_{1\leq m\leq M} $ be the $n$-th column vector of precoding matrix ${{\bf{W}}}$, the Fourier transform of ${{\bf{w}}_n}$ is
\begin{equation}
\setlength{\abovedisplayskip}{2.5pt}
\setlength{\belowdisplayskip}{2.5pt}
{W_n}(\omega ) = \sum\limits_{m = 1}^M {w_{n,m}{e^{ - {\rm{j}}\left( {m - 1} \right)\omega }}}.
\end{equation}
Thus, ${W_n}(\omega ){W_k^*}(\omega )$ can be written as
\begin{equation}
\small
\setlength{\abovedisplayskip}{1.5pt}
\setlength{\belowdisplayskip}{1.5pt}
\begin{array}{l}
\begin{aligned}
&\!\!\!\!\!\!\!\!\!\!\!{W_n}(\omega ){W_k^*}(\omega )
 =\sum\limits_{m = 1}^{{M}} { {\sum\limits_{l = 1}^{{M}} {w_{n,m}w_{k,l}^*{e^{ -{\rm{j}}\left( {m - l} \right)\omega }}}
  } } \\
 &\!\!\!\!\!\!\!\!\!\!\! \overset{\left(a\right)}{=} \!\! \sum\limits_{a = 0}^{{M} \!- \!1} {{\sum\limits_{m = 1}^{{M}\! - \!a} \!\!{w_{n,m}w_{k,m+a}^*{e^{  {\rm{j}}a\omega }} } }} 
 \!\!+\!\! \sum\limits_{a =\! 1 \!-\! {M\!}}^{ - 1} {{\sum\limits_{m\! = \!1 - \!a}^{{M}}  {w_{n,m}w_{k,m+a}^*{e^{  {\rm{j}}a\omega }} } } }  \\
   &\!\!\!\!\!\!\!\!\!\!\! \overset{\left(b\right)}{=}\!\! \sum\limits_{a = 1-M}^{{M} - 1} {{{R _{{{\bf{w}}_{n}}{{\bf{w}}_{k}}}}\left( {a} \right){e^{{\rm{j}}a\omega}}} }
\end{aligned}
\end{array}
\end{equation}
where (a) follows by letting $a=l-m$  , and (b) is due to the definition of AACF and ACCF in \eqref{acf}. Letting $\left\{{\bf{w}}_1,{\bf{w}}_2\right\}$ and $\left\{{\bf{w}}_3,{\bf{w}}_4\right\}$ be mutually orthogonal complementary, with the properties \eqref{occ} of OCC, it is easy to know that
\setlength{\abovedisplayskip}{2pt}
\setlength{\belowdisplayskip}{2pt}
{\small
\[{\sum\limits_{n = 1}^4 \!{|{W_n}(\omega ){|^2}\! =\! \sum\limits_{a =\! 1 -\! M}^{M \!- \!1} {\sum\limits_{n = 1}^4 {{R _{{{\bf{w}}_n}}}\!\left( \!a \right)\!{e^{ \! {\rm{j}}a\omega }}} }  \!=\! \sum\limits_{n = 1}^4 {{R _{{{\bf{w}}_n}}}\!\left( 0 \right)} } },\]
\[{\sum\limits_{n = 1}^2\! {{W_n}\!(\omega ){W_{n + 2}^*}(\omega )} \! = \!\sum\limits_{a = 1 \!- \!M}^{M\! - \!1} \!{\sum\limits_{n = 1}^2 \!{{R _{{{\bf{w}}_n}{{\bf{w}}_{n + 2}}}}\!\left( \!a \right)\!{e^{ {\rm{j}}a\omega }}} }  = 0}.
\]}
Therefore, \eqref{sum} can be rewritten as 
\begin{equation}
\small
\setlength{\abovedisplayskip}{2pt}
\setlength{\belowdisplayskip}{2pt}
\sum\limits_{t = 1}^4 {|{S_t}(\omega ){|^2}}   = \alpha \sum\limits_{n = 1}^4 {{R _{{{\bf{w}}_n}}}\left( 0 \right)}  = \alpha  \cdot \rm{tr}\left( {{\bf{W}}{{\bf{W}}^H}} \right) = 4\alpha
\end{equation}
which { is constant for all $\theta \in (-\pi,\pi]$}, and proves the theorem.
\end{document}